\documentclass[12pt,prd,superscriptaddress,preprintnumbers,amsmath,amssymb,nofootinbib]{revtex4-2}

\usepackage{xcolor}
\usepackage{graphicx}
\usepackage{tabularx}
\usepackage{latexsym}
\usepackage{float}
\usepackage{epsfig}
\usepackage{longtable}
\usepackage{orcidlink}
\usepackage[outdir=./]{epstopdf}
\usepackage{verbatim}
\usepackage{comment}
\usepackage[compat=1.1.0]{tikz-feynman}
\usetikzlibrary{decorations.markings}
\usetikzlibrary{positioning, arrows.meta}
\allowdisplaybreaks

\begin{document}
\title{Probing Signatures of Right-Handed Neutrinos via $b \to s \nu \bar\nu$ Decays}

\author{Prisha \orcidlink{0009-0009-5521-7761}}
\email{p23ph0008@iitj.ac.in}
\affiliation{Department of Physics, IIT Jodhpur, India}

\author{Priyanka Boora \orcidlink{0009-0003-4248-375X}}
\email{2020rpy9601@mnit.ac.in}
\affiliation{Department of Physics, MNIT Jaipur, India}

\author{Dinesh Kumar \orcidlink{0000-0001-6884-3790}}
\email{dinesh@uniraj.ac.in}
\affiliation{Department of Physics, University of Rajasthan, Jaipur, India}

\author{Jitendra Kumar \orcidlink{0000-0002-8465-433X}}
\email{jitendra@iitj.ac.in}
\affiliation{Department of Physics, IIT Jodhpur, India}

\begin{abstract}
The recent Belle II measurement of $\mathcal{B}(B^+\to K^+\nu\bar{\nu})$, which deviates from the Standard Model prediction, provides a strong motivation to search for New Physics in $b\to s\nu\bar{\nu}$ transitions. We perform a model-independent analysis within the low-energy effective theory, focusing on dimension-six vector operators involving right-handed neutrinos. Using the Belle II measurement of $\mathcal{B}({B\to K\nu\bar{\nu}})$ together with the upper limit on $\mathcal{B}({B\to K^{*}\nu\bar{\nu}})$, we constrain the new physics interactions. We study the correlation of $\mathcal{B}({B\to K\nu\bar{\nu}})$ with $\mathcal{B}({B\to K^{*}\nu\bar{\nu}})$, $\mathcal{B}({B_s\to\phi\nu\bar{\nu}})$, $\mathcal{B}({\Lambda_b\to\Lambda\nu\bar{\nu}})$, and the longitudinal polarization fraction $F_L^{K^*}$. We find sizeable enhancements in all branching fractions studied, while $F_L^{K^*}$ offers complementary sensitivity to the underlying RHN interaction structure. These results provide testable signatures of right-handed neutrino interactions at Belle~II and LHCb, and at the future FCC-ee experiment.

\end{abstract}
\maketitle
\newpage
\section{Introduction}  \label{sec:introduction} 
The Standard Model (SM) of particle physics has been remarkably successful in describing the fundamental particles and their interactions. Despite its impressive agreement with a wide range of experimental observations, several unresolved issues and emerging tensions between theoretical predictions and experimental measurements indicate that the SM may not be the complete theory of nature. These discrepancies motivate the exploration of possible new physics (NP) scenarios beyond the SM. In this context, flavor physics, which studies transitions among different generations of quarks and leptons, provides a highly sensitive framework for probing possible NP effects~\cite{Albrecht:2021tul,Crivellin:2016ivz, Botella:2012ju, Straub:2011gt, Buras:2009if}. 

Among the various processes in flavor physics, rare flavor-changing transitions are particularly interesting because their suppressed nature within the SM makes them highly sensitive to possible contributions from heavy virtual particles arising in extended theories. In particular, flavor-changing neutral current (FCNC) processes play a central role in the search for NP, since they are strongly suppressed through the Glashow--Iliopoulos--Maiani (GIM) mechanism~\cite{Glashow:1970gm}. Such transitions are forbidden at the tree level and occur only through higher-order loop diagrams, leading to highly suppressed decay rates within the SM. Consequently, even small deviations from the SM predictions in branching fractions and other observables can provide important hints of possible NP contributions~\cite{Buras:2014fpa, Bause:2021cna, Browder:2021hbl, Allwicher:2023xba}.

Rare $B$-meson decays mediated through FCNC transitions have therefore emerged as powerful tools for testing the flavor structure of the SM~\cite{Buchalla:1995vs, Colangelo:1996ay, Altmannshofer:2009ma, Descotes-Genon:2020buf}. In recent years, several measurements in rare $B$-decays have shown intriguing tensions with the SM expectations in branching fractions, angular observables, CP-asymmetry measurements, and tests of lepton flavor universality~\cite{LHCb:2022qnv, BELLE:2019xld, BaBar:2012mrf, Belle:2016fev, LHCb:2021lvy, LHCb:2021trn, LHCb:2019efc, LHCb:2019hip, LHCb:2017avl, LHCb:2014vgu}. These observed deviations have further strengthened the motivation for detailed phenomenological studies of rare FCNC-mediated $B$ decays.

Semileptonic $B$ decays mediated through FCNC transitions provide particularly important probes for testing the SM and exploring possible NP effects~\cite{LHCb:2022qnv, BELLE:2019xld, BaBar:2012mrf, Belle:2016fev, LHCb:2021lvy, LHCb:2021trn, LHCb:2019efc, LHCb:2019hip, LHCb:2017avl, LHCb:2014vgu}. Although considerable attention has been devoted to the charged-lepton modes associated with the $b \to s \ell^{+}\ell^{-}$ transition due to the observed anomalies in several observables, but the neutrino modes mediated through $b \to s \nu \bar{\nu}$ transitions also offer a complementary and theoretically cleaner framework for investigating the flavor sector. In particular, processes such as $B \to K \nu \bar{\nu}$ and $B \to K^{*} \nu \bar{\nu}$ are theoretically very clean because the neutrinos in the final state eliminate sizable long-distance electromagnetic effects and significantly reduce hadronic uncertainties~\cite{Allwicher:2023xba, Fael:2025xmi, Hu:2025zua, Das:2025zrn, Rosauro-Alcaraz:2024mvx, Altmannshofer:2025eor, Altmannshofer:2021qrr, Buras:2024mnq, Felkl:2021uxi}. Consequently, the corresponding SM predictions are comparatively precise and reliable. At the same time, these decays remain highly sensitive to NP effects from scenarios involving additional gauge bosons, leptoquarks, supersymmetry, and non-standard neutrino interactions~\cite{Becirevic:2024iyi, Buras:2014fpa, Buras:2024mnq, Browder:2021hbl, He:2021yoz, Crivellin:2025qsq, Chen:2025npb, Altmannshofer:2023hkn, Wang:2023trd, Aliev:2025hyp}.

Experimentally, the study of $b \to s \nu \bar{\nu}$ transitions is highly challenging due to the presence of neutrinos in the final state, which escape detection and lead to missing-energy signatures~\cite{Belle-II:2023esi}. Crucially, any light right-handed neutrinos (RHNs) or other weakly interacting neutral particles would produce identical missing-energy signatures, making these decay modes particularly sensitive probes of the invisible particle spectrum beyond the SM. Nevertheless, significant progress has been made in recent years through dedicated searches at $B$-factory experiments such as BaBar \cite{BaBar:2012mrf}, Belle \cite{Belle:2017oht}, and Belle II~\cite{Belle-II:2023esi}. The high luminosity and improved detector capabilities at Belle II are expected to considerably enhance the sensitivity to these rare decay channels and provide stringent tests of the SM in the flavor sector~\cite{Belle-II:2018jsg}. 

\begin{table}[ht]
\centering
\begin{tabular}{lcc}
\hline\hline
Observable & Value & Experiment/Theory \\
\hline
$\mathcal{B}(B^{+}\to K^{+}\nu\bar{\nu})_{\rm had}$ &
$\left(1.1^{+0.9}_{-0.8}\,{\rm(stat)}^{+0.8}_{-0.5}\,{\rm(syst)}\right)\times10^{-5}$ &
Belle II~\cite{Belle-II:2023esi} \\

$\mathcal{B}(B^{+}\to K^{+}\nu\bar{\nu})_{\rm inc}$ &
$\left(2.7 \pm 0.5\,{\rm(stat)} \pm 0.5\,{\rm(syst)}\right)\times10^{-5}$ &
Belle II~\cite{Belle-II:2023esi} \\

$\mathcal{B}(B\to K^{+}\nu\bar{\nu})_{\rm combined}$ &
$(2.3 \pm 0.7)\times10^{-5}$ &
Belle II~\cite{Belle-II:2023esi} \\

$\mathcal{B}(B\to K^{*}\nu\bar{\nu})$ &
$< 2.7\times10^{-5}$ &
Belle~\cite{Belle:2017oht} \\
\hline
$\mathcal{B}(B\to K\nu\bar{\nu})_{\rm SM}$ &
$(4.6 \pm 0.5)\times10^{-6}$ &
SM~\cite{Browder:2021hbl} \\

$\mathcal{B}(B\to K^{*}\nu\bar{\nu})_{\rm SM}$ &
$(9.2 \pm 0.7)\times10^{-6}$ &
SM~\cite{Parrott:2022zte} \\
\hline\hline
\end{tabular}
\caption{Experimental measurements and SM predictions for the $B \to K^{(*)}\nu\bar{\nu}$ decay modes.}
\label{tab:BKnn_exp}
\end{table}

Recently, the Belle II Collaboration reported the first evidence for the $B^{+} \to K^{+} \nu \bar{\nu}$ decay using both hadronic-tagging and inclusive reconstruction methods~\cite{Belle-II:2023esi}. The corresponding experimental measurements, along with the SM predictions for the $B \to K^{(*)}\nu\bar{\nu}$ decay modes, are summarized in table~\ref{tab:BKnn_exp}. Combining the hadronic-tagging and inclusive analyses, Belle II measured a branching fraction for $B \to K \nu \bar{\nu}$ that lies approximately $2.7\sigma$ above the corresponding SM prediction. This excess, if confirmed, would require enhanced short-distance contributions beyond the SM, such as new vector or axial-vector currents involving both left- and right-handed neutrinos. On the other hand, for the $B \to K^{*}\nu\bar{\nu}$ channel, no precise measurement is currently available, although the Belle Collaboration has reported an upper limit at $90\%$ confidence level~\cite{Belle:2017oht}. The $B \to K^{*}$ mode offers complementary sensitivity through additional angular observables and distinct Lorentz structures, making it a valuable channel for disentangling different NP contributions. These measurements and limits continue to motivate detailed phenomenological investigations of the $b \to s \nu \bar{\nu}$ transition as a sensitive probe of physics beyond the SM.

Owing to the presence of neutrinos in the final state, these decay modes provide an ideal framework for exploring possible NP effects associated with the neutrino sector. 
In particular, scenarios involving right-handed neutrinos have attracted considerable attention in the phenomenological study of FCNC processes~\cite{Becirevic:2024iyi, Felkl:2023ayn, Rosauro-Alcaraz:2024mvx}. The RHNs arise naturally in several extensions of the SM that attempt to explain the origin of neutrino masses and the observed neutrino oscillation phenomena, including type-I seesaw, inverse seesaw, and related low-scale neutrino mass models~\cite{Escalona:2025jla, Rosauro-Alcaraz:2024mvx, DelleRose:2019ukt}.
Since light RHNs would be invisible in the detector and contribute to the missing-energy signature in the same way as SM neutrinos, the $B \to K^{(*)} \nu \bar{\nu}$ modes are uniquely suited to probing such scenarios without requiring a dedicated new-physics signature. The inclusion of RHNs can modify the structure of the effective interactions and lead to observable effects in branching fractions and other observables associated with these decays. Such effects can be systematically analyzed within the framework of the low-energy effective theory (LEFT), which allows the inclusion of both left-handed and right-handed neutrino operators with different Lorentz structures, including scalar, vector, and tensor currents.

Motivated by the recent Belle~II results for the rare decay $B^+\to K^+\nu\bar{\nu}$, we perform a model-independent analysis of $b\to s\nu\bar{\nu}$ transitions within the 
framework of the low-energy effective theory, focusing on dimension-six vector operators involving right-handed neutrinos. Extending the standard operator basis to include 
right-handed neutrino interactions $b \to s \nu_R \bar{\nu}_R$, we constrain the Wilson coefficients (WCs) $C_V^{LR}$ and $C_V^{RR}$ using the Belle~II measurement of $\mathcal{B}(B^+\to K^+\nu\bar{\nu})$ together with the current Belle upper limit on $\mathcal{B}(B\to K^{*}\nu\bar{\nu})$, thereby determining the allowed parameter space and providing a basis for predicting observables in related decay modes. We then investigate their impact on the branching 
fractions of $B\to K\nu\bar{\nu}$, $B\to K^{*}\nu\bar{\nu}$, $B_s\to\phi\nu\bar{\nu}$, 
and $\Lambda_b\to\Lambda\nu\bar{\nu}$, as well as on the longitudinal polarization 
fraction $F_L^{K^*}$.

The paper is organized as follows. In Section~\ref{sec:framework}, we present the theoretical framework and discuss the effective Hamiltonian for the $b \to s \nu \bar{\nu}$ transition in the presence of RHNs within LEFT. In Section~\ref{sec:observables}, we discuss the relevant observables and outline the methodology used in our analysis. The numerical analysis and phenomenological results are presented in Section~\ref{sec:results}. Finally, we summarize our findings and conclude in Section~\ref{sec:conclusion}.

\section{Theoretical Framework} \label{sec:framework}
The $b \to s \nu\bar{\nu}$ transition proceeds in the SM exclusively through loop-level Z-penguin and box diagrams, with the top quark dominating the loop function due to its large mass. At energies well below the electroweak scale, the heavy degrees of freedom, the $W$ and $Z$ bosons, the top quark, and the Higgs, are integrated out, and their effects are encoded into WCs of local four-fermion operators. The resulting effective Hamiltonian within  the SM is

\begin{equation}
\mathcal{H}_{\rm eff}^{\rm SM} \;=\; -\frac{4G_F}{\sqrt{2}}\, \frac{\alpha_{EM}}{4 \pi}
V_{ts}^{*}V_{tb}\,\mathcal{C}_{L}^{\rm SM}\,\mathcal{O}_{L} 
\;+\; \text{h.c.},
\label{eq:Heff_SM}
\end{equation}

with Fermi constant $G_F = 1.1663788 \times 10^{-5}~\text{GeV}^{-2}$~\cite{PhysRevD.110.030001} and relevant $V_{ts}^{*}V_{tb}$ product of Cabibbo--Kobayashi--Maskawa (CKM) matrix elements governing the $b \to s$ flavor transition. The only dimension-six vector operator contributing to the $b \to s \nu\nu$ transition within the SM is given by

\begin{equation}
\mathcal{O}_{L} \;=\; 
\left(\bar{s}\,\gamma_{\mu}P_L\, b\right)
\left(\bar{\nu}\,\gamma^{\mu}P_L\,\nu\right),
\label{eq:OL}
\end{equation}

The corresponding Wilson coefficient is given by
\begin{equation}
\mathcal{C}_{L}^{\rm SM} \;=\; -\frac{2X_t}{s_w^2},
\label{eq:CSM}
\end{equation}
where $s_w \equiv \sin\theta_W$ and $X_t$ encodes the top-quark loop function. Including next-to-leading-order electroweak corrections, the numerical value is $X_t/s_w^2 = 6.352 \pm 0.074$~\cite{Buras:2014fpa}, giving $\mathcal{C}_{L}^{\rm SM} \approx -12.704$. This coefficient is real, and its precise value is one of the key inputs in the theoretical prediction for the $b \to s \nu\bar{\nu}$ branching fractions.

To explore NP contributions in a model-independent way, we extend the SM operator basis by including dimension-6 four-fermion RHN operators consistent with the unbroken $SU(3)_C \times U(1)_{\rm EM}$ symmetry 
of the LEFT. In the present work, we focus on vector operators, which constitute the dominant contributions considered in many studies of $b\to s\nu\bar{\nu}$ transitions with RHNs and provide a simple framework for investigating the impact of right-handed neutrino interactions on rare $B$ decays. The relevant operators are of the form $(\bar{s}\gamma_\mu P_X b)(\bar{\nu}\gamma^\mu P_Y\nu)$ with $X,Y\in{L,R}$, and are given by
\begin{align}
\mathcal{O}_{V}^{LR} &\;=\; 
\left(\bar{s}\,\gamma_{\mu}P_L\, b\right)
\left(\bar{\nu}\,\gamma^{\mu}P_R\,\nu\right), 
\label{eq:OLR} \\
\mathcal{O}_{V}^{RR} &\;=\; 
\left(\bar{s}\,\gamma_{\mu}P_R\, b\right)
\left(\bar{\nu}\,\gamma^{\mu}P_R\,\nu\right).
\label{eq:ORR}
\end{align}

The operators $\mathcal{O}_V^{LR}$ and $\mathcal{O}_V^{RR}$, involve right-handed neutrino currents $(\bar{\nu}\, \gamma^\mu P_R\,\nu)$ and are therefore generated only in theories that contain RHNs as explicit degrees of freedom, such as low-scale seesaw models~\cite{King:2025eqv} or $Z'$ and leptoquark scenarios with right-handed neutrino couplings \cite{Browder:2021hbl}.

The general effective Hamiltonian for the $b \to s\nu\bar{\nu}$ transition, including both four-fermion NP operators, can be written as
\begin{equation}
\mathcal{H}_{\rm eff} \;=\; -\frac{4G_F}{\sqrt{2}}\, \frac{\alpha_{EM}}{4 \pi} V_{ts}^{*}V_{tb}
\left(
\mathcal{C}_{L}^{\rm SM}\,\mathcal{O}_{L} 
\;+\; \mathcal{C}_{V}^{LR}\,\mathcal{O}_{V}^{LR}
\;+\; \mathcal{C}_{V}^{RR}\,\mathcal{O}_{V}^{RR}
\right) + \text{h.c.},
\label{eq:Heff_NP}
\end{equation}

where $\mathcal{C}_{V}^{LR}$ and $\mathcal{C}_{V}^{RR}$ are the NP WCs that determine the strength of the corresponding NP operator.

Since the experimental observables correspond to branching fractions summed over all neutrino flavors, we assume lepton flavor universal (LFU) new-physics interactions, taking the WCs to be identical for all neutrino generations, 
$\mathcal{C}_i^{ee} = \mathcal{C}_i^{\mu\mu} = \mathcal{C}_i^{\tau\tau} 
\equiv \mathcal{C}_i$, so that the new-physics contribution affects all three neutrino flavors equally. Under this assumption, the WCs $\mathcal{C}_V^{LR}$ and $\mathcal{C}_V^{RR}$ are constrained using the Belle~II measurement of $\mathcal{B}(B^+\to K^+\nu\bar{\nu})$~\cite{Belle-II:2023esi} and the current Belle upper limit on 
$\mathcal{B}(B\to K^*\nu\bar{\nu})$~\cite{Belle:2017oht}.

\section{Observables} \label{sec:observables}
The rare decays $B\to K^{(*)}\nu\bar{\nu}, B_s\to\phi\nu\bar{\nu}$, and $\Lambda_b\to\Lambda\nu\bar{\nu}$ provide complementary probes of the underlying $b\to s\nu\bar{\nu}$ transition. Their branching fractions are sensitive to the overall strength of the effective operators introduced in section~\ref{sec:framework} and therefore offer direct constraints on the corresponding WCs. In addition, the longitudinal polarization fraction $F_L^{K^*}$ in $B\to K^*\nu\bar{\nu}$ serves as an important angular observable that probes the helicity structure of the underlying interaction and distinguishes different new-physics scenarios.

In the presence of right-handed neutrino operators, these observables receive characteristic modifications through different combinations of the WCs $C_{V}^{LR}$ and $C_{V}^{RR}$. While the pseudoscalar final state is primarily sensitive to the combination $C_V^{LR}+C_V^{RR}$, vector final states such as $K^*$ and $\phi$ allow separate sensitivity to both the sum and difference of these coefficients through their longitudinal and transverse helicity amplitudes. The baryonic decay $\Lambda_b\to\Lambda\nu\bar{\nu}$ provides an additional independent probe with a distinct hadronic structure. In the following subsections, we present the differential decay rates and observables used in our numerical analysis.
\subsection{$B \to K \nu \bar{\nu}$ Decay} \label{BK}

The general differential decay width for $B \to K\nu\bar{\nu}$, including all dimension-six RHN vector operators and summing over all lepton generations, is given by

\begin{equation}
\begin{split}
\frac{d\Gamma}{dq^2}(B \to K\nu\bar{\nu}) = \,&
\frac{G_F^2\,|V_{tb}V_{ts}^*|^2\,\alpha_{\rm EM}^2}
{192 \times 16\pi^5 m_B^3}\,q^2\,\lambda_K^{1/2}(q^2)\,
\left(\mathcal{H}_V^s\right)^2 \sum_{\alpha,\beta}\left[
\left|\mathcal{C}_L^{\rm SM}\delta_{\alpha\beta}\right|^2
+ \left|[\mathcal{C}_V^{LR}]^{\alpha\beta}
+ [\mathcal{C}_V^{RR}]^{\alpha\beta}\right|^2
\right]
\end{split}
\label{eq:dGamma_K}
\end{equation}
where the indices $\alpha,\beta \in \{e,\mu,\tau\}$ denote the neutrino flavors and $\lambda_K(q^2) \equiv \lambda(m_B^2, m_K^2, q^2)$ is the K\"all\'en function and the hadronic helicity amplitude is
\begin{equation}
\mathcal{H}_V^s(q^2) = \sqrt{\frac{\lambda_K(q^2)}{q^2}}\,f_+(q^2),
\label{eq:HVs}
\end{equation}
with $f_+(q^2)$ the $B\to K$ vector form factor taken from lattice QCD~\cite{Bailey:2015dka}. 
\subsection{$B \to K^* \nu \bar{\nu}$ Decay} \label{BKstr}

The general differential decay width for $B \to K^*\nu\bar{\nu}$ receives contributions from three helicity amplitudes and is given by

\begin{equation}
\begin{aligned}
\frac{d \Gamma}{d q^2}\left(B \rightarrow K^* \bar{\nu} \nu\right)= & \frac{G_F^2\left|V_{t b} V_{t s}^*\right|^2 \alpha_{\mathrm{EM}}^2}{192 \times 16 \pi^5 m_B^3} q^2 \lambda_{K^*}^{1 / 2}\left(q^2\right) \times \\
 & \sum_{\alpha, \beta} \left|\mathcal{C}_{L}^{\mathrm{SM}} \delta_{\alpha \beta}\right|^2\left(\mathcal{H}_{V,+}^2+\mathcal{H}_{V,-}^2 + \mathcal{H}_{V, 0}^2\right) \\
& + \left(\left|\left[\mathcal{C}^{L R}_V\right]^{\alpha \beta}\right|^2+\left|\left[\mathcal{C}^{R R}_V\right]^{\alpha \beta}\right|^2\right)\left(\mathcal{H}_{V,+}^2+\mathcal{H}_{V,-}^2\right) \\
& +\left|\left[\mathcal{C}^{L R}_V\right]^{\alpha \beta}-\left[C^{R R}_V\right]^{\alpha \beta}\right|^2 \mathcal{H}_{V, 0}^2-4 \mathcal{R}\left[\left[\mathcal{C}^{L R}_V\right]^{\alpha \beta}\left[\mathcal{C}^{R R \ast}_{V}\right]^{\alpha \beta}\right] \mathcal{H}_{V,+} \mathcal{H}_{V,-}
\label{eq:dGamma_Kst}
\end{aligned}
\end{equation}

where $\lambda_{K^*}(q^2) \equiv \lambda(m_B^2,m_{K^*}^2,q^2)$ is defined analogously to $\lambda_K$. The helicity amplitudes are
\begin{align}
\mathcal{H}_{V,\pm}(q^2) &= (m_B + m_{K^*})\,A_1(q^2) 
\mp \frac{\sqrt{\lambda_{K^*}(q^2)}}{m_B + m_{K^*}}\,V(q^2), 
\label{eq:HVpm}\\
\mathcal{H}_{V,0}(q^2) &= \frac{8m_B m_{K^*}}{\sqrt{q^2}}\,A_{12}(q^2)
\label{eq:HV0}
\end{align}

with the form factors $V$, $A_1$, and $A_{12}$ taken from LCSR calculations~\cite{Bharucha:2015bzk}. 

Besides the branching fraction, the longitudinal polarization fraction of the $K^*$ meson provides an additional probe of the chiral structure of the effective interaction. Since different operator combinations contribute differently to the longitudinal and transverse helicity amplitudes, $F_L^{K^*}$ can help discriminate between new-physics scenarios that predict similar branching fractions. The integrated longitudinal polarization fraction is defined as

Whereas the longitudinal polarization fraction $F_L(q^2) = (d\Gamma_L/dq^2)/(d\Gamma/dq^2)$, with $d\Gamma_L/dq^2$ obtained from equation~\ref{eq:dGamma_Kst} by retaining only 
the $H_{V,0}^2$ terms, integrates to

\begin{equation}
\langle F_L \rangle = \frac{\displaystyle\int_{q^2_{\min}}^{(m_B-m_{K^*})^2} 
\frac{d\Gamma_L}{dq^2}\,dq^2}{\displaystyle\int_{q^2_{\min}}^{(m_B-m_{K^*})^2} 
\frac{d\Gamma}{dq^2}\,dq^2},
\label{eq:FL_int}
\end{equation}

\subsection{$B_s \to \phi \nu\bar{\nu}$ Decay} \label{Bphi}

The $B_s \to \phi\nu\bar{\nu}$ decay proceeds through the same $b \to s\nu\bar{\nu}$ transition and has an identical helicity structure to $B \to K^*\nu\bar{\nu}$. The differential decay width is obtained from equation~\ref{eq:dGamma_Kst} by the  replacements $m_B \to m_{B_s}$, $m_{K^*} \to m_\phi$, $\tau_B \to \tau_{B_s}$, and $\lambda_{K^*} \to \lambda_\phi(q^2) \equiv \lambda(m_{B_s}^2, m_\phi^2, q^2)$, 
with the helicity amplitudes defined as

\begin{align}
\mathcal{H}_{V,\pm}(q^2) &= (m_{B_s} + m_\phi)\,A_1(q^2) 
\mp \frac{\sqrt{\lambda_\phi(q^2)}}{m_{B_s} + m_\phi}\,V(q^2), 
\label{eq:HVpm_phi}\\
\mathcal{H}_{V,0}(q^2) &= \frac{8\,m_{B_s}\,m_\phi}{\sqrt{q^2}}\,
A_{12}(q^2)
\label{eq:HV0_phi}
\end{align}

where the $B_s \to \phi$ form factors $V$, $A_1$, and $A_{12}$ are taken from LCSR calculations \cite{Bharucha:2015bzk}, parametrized in the $z$-expansion.

\subsection{$\Lambda_b \to \Lambda \nu\bar{\nu}$ Decay} \label{Lambda}

The $\Lambda_b \to \Lambda\nu\bar{\nu}$ decay proceeds through the same $b \to s\nu\bar{\nu}$ transition but involves a baryon in the initial and final state, leading to a distinct and complex helicity structure compared to the meson modes above. The differential branching fraction is

\begin{equation}
\frac{d \Gamma}{dq^2}(\Lambda_b \to \Lambda\nu\bar{\nu}) =
\frac{G_F^2\,|V_{tb}V_{ts}^*|^2\,
\alpha_{\rm EM}^2}{3 \times 2^{11}\pi^5\,m_{\Lambda_b}^3}\,
\sqrt{\lambda_\Lambda(q^2)}\sum_{\alpha, \beta}\left[
\mathcal{C}_-^{\alpha \beta}\,\mathcal{F}_A(q^2)
+ \mathcal{C}_+^{\alpha \beta}\,\mathcal{F}_V(q^2)\right]
\label{eq:dBR_Lambda}
\end{equation}

with 
\begin{align}
\mathcal{C}_-^{\alpha \beta} &= 
\left|\mathcal{C}_V^{LR,\alpha \beta} - 
\mathcal{C}_V^{RR,\alpha \beta}\right|^2, 
\label{eq:Cminus}\\
\mathcal{C}_+^{\alpha \beta} &= 
\left|\mathcal{C}_V^{LR,\alpha \beta} + 
\mathcal{C}_V^{RR,\alpha \beta}\right|^2
\label{eq:Cplus}
\end{align}

where $\lambda_\Lambda(q^2) = s_+(q^2)\,s_-(q^2)$ with $s_\pm(q^2) = (m_{\Lambda_b} \pm m_\Lambda)^2 - q^2$, and the hadronic structure functions are

\begin{align}
\mathcal{F}_A(q^2) &= (m_{\Lambda_b} - m_\Lambda)^2\,
s_+(q^2)\,f_0^A(q^2)^2 
+ 2q^2\,s_+(q^2)\,f_\perp^A(q^2)^2, 
\label{eq:FA}\\
\mathcal{F}_V(q^2) &= (m_{\Lambda_b} + m_\Lambda)^2\,
s_-(q^2)\,f_0^V(q^2)^2 
+ 2q^2\,s_-(q^2)\,f_\perp^V(q^2)^2
\label{eq:FV}
\end{align}

where $f_+^{V,A}$ and $f_\perp^{V,A}$ are the vector and axial $\Lambda_b \to \Lambda$ form factors taken from lattice QCD~\cite{Detmold:2016pkz}. 
\section{Results}  \label{sec:results}
\begin{figure}[h]
\centering
\includegraphics[width=0.60\textwidth]{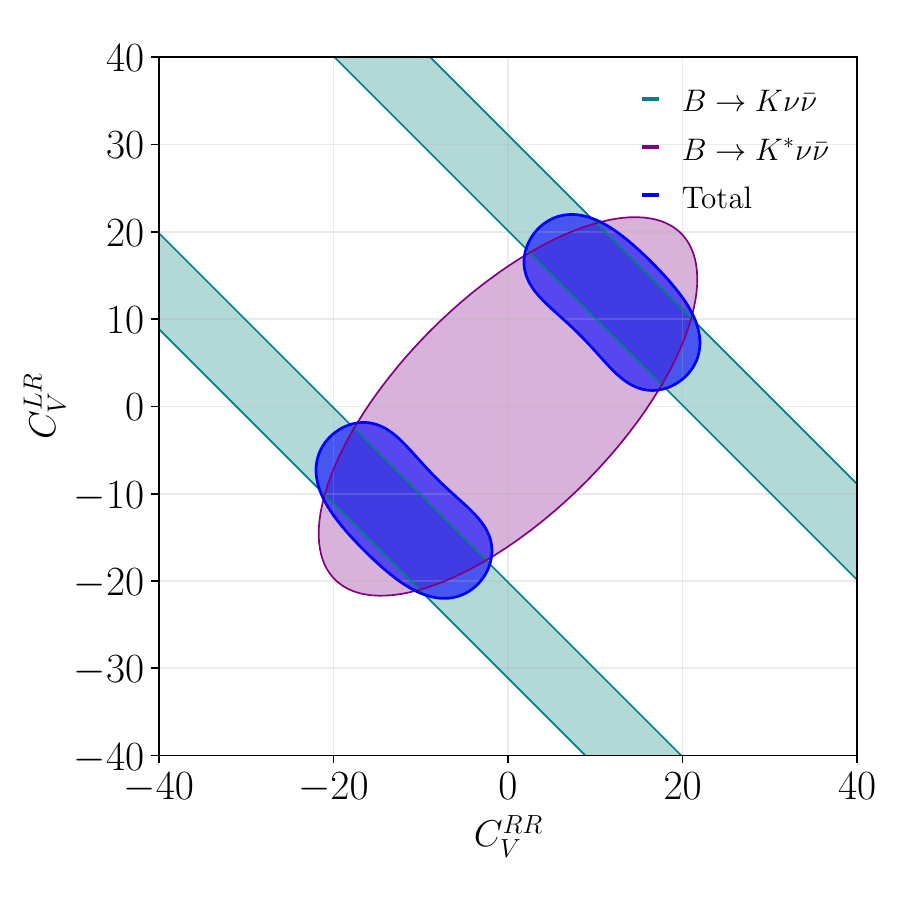}
\caption{Constraints on the WCs 
$(\mathcal{C}_V^{RR},\,\mathcal{C}_V^{LR})$ in the RHN scenario from $B\to K\nu\bar{\nu}$ (teal) and $B\to K^*\nu\bar{\nu}$ (purple) at $1\sigma$. The blue shaded regions show the combined allowed parameter space, with the SM corresponding to the origin $\mathcal{C}_V^{RR} = \mathcal{C}_V^{LR} = 0$.}
\label{wc_plot}
\end{figure}

To identify the regions of WC space consistent with current data, we perform a $\chi^2$ minimization over the $(\mathcal{C}_V^{LR},\,\mathcal{C}_V^{RR})$ plane, constructing separate $\chi^2$ functions for $B\to K\nu\bar{\nu}$ and $B\to K^*\nu\bar{\nu}$ and combining them into a total $\chi^2_{\rm tot} = \chi^2_{B\to K\nu\bar{\nu}} + \chi^2_{B\to K^*\nu\bar{\nu}}$. The individual $1\sigma$ allowed regions for each mode are defined by $\Delta\chi^2 \leq 1$ with respect to the minimum of the corresponding individual $\chi^2$ function, while the combined $1\sigma$ region is determined by $\Delta\chi^2_{\rm tot} \leq 2.30$, with the results shown in figure~\ref{wc_plot}. As seen from equation~\ref{eq:dGamma_K}, $\mathcal{B}(B\to K\nu\bar{\nu})$ depends only on $|\mathcal{C}_V^{LR} + \mathcal{C}_V^{RR}|^2$, producing two diagonal bands of negative slope in the $(\mathcal{C}_V^{LR},\,\mathcal{C}_V^{RR})$ plane. The $B\to K^*\nu\bar{\nu}$ constraint is more complex, as from equation~\ref{eq:dGamma_Kst}, the transverse helicity amplitudes depend on $|\mathcal{C}_V^{LR}|^2 + |\mathcal{C}_V^{RR}|^2$, the longitudinal amplitude on $|\mathcal{C}_V^{LR} - \mathcal{C}_V^{RR}|^2$, and an additional interference term proportional to $\mathcal{R}\big[\mathcal{C}_V^{LR}\,\mathcal{C}_V^{RR*}\big]$ also contributes, together producing elliptical allowed regions oriented along both diagonals and complementary to the $B\to K\nu\bar{\nu}$ bands. The overlap of the two constraints yields two distinct allowed islands 
in the $(\mathcal{C}_V^{LR},\,\mathcal{C}_V^{RR})$ plane, highlighted in blue in figure~\ref{wc_plot}, identifying the regions where right-handed neutrino contributions are consistent with current 
experimental data. Notably, the SM point 
$\mathcal{C}_V^{LR} = \mathcal{C}_V^{RR} = 0$ lies outside both allowed islands, indicating that non-zero right-handed neutrino contributions are favored by the combined experimental constraints 
at $1\sigma$.

\begin{figure}[!htb]
\centering
\includegraphics[width=0.49\textwidth]{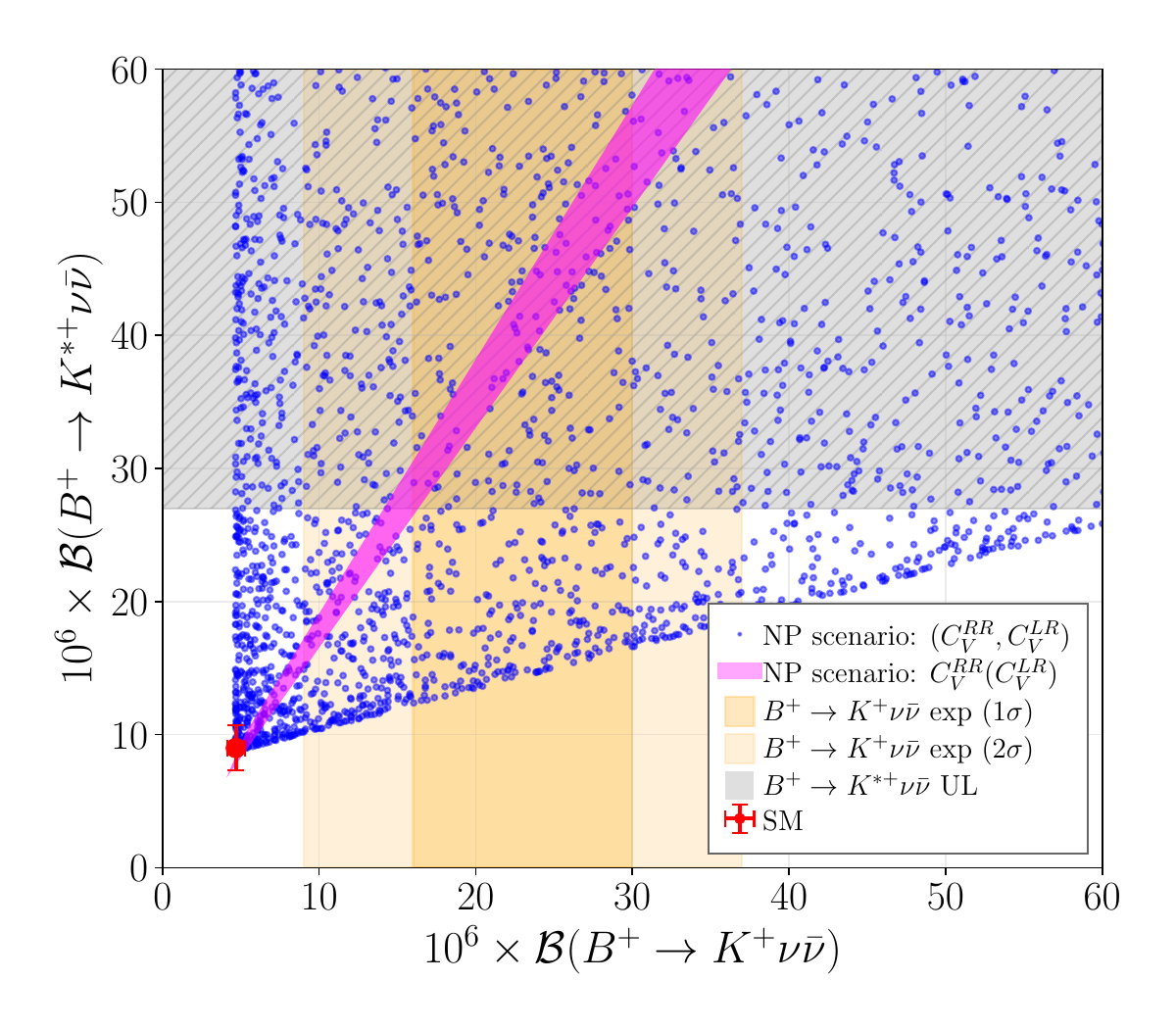}
\includegraphics[width=0.49\textwidth]{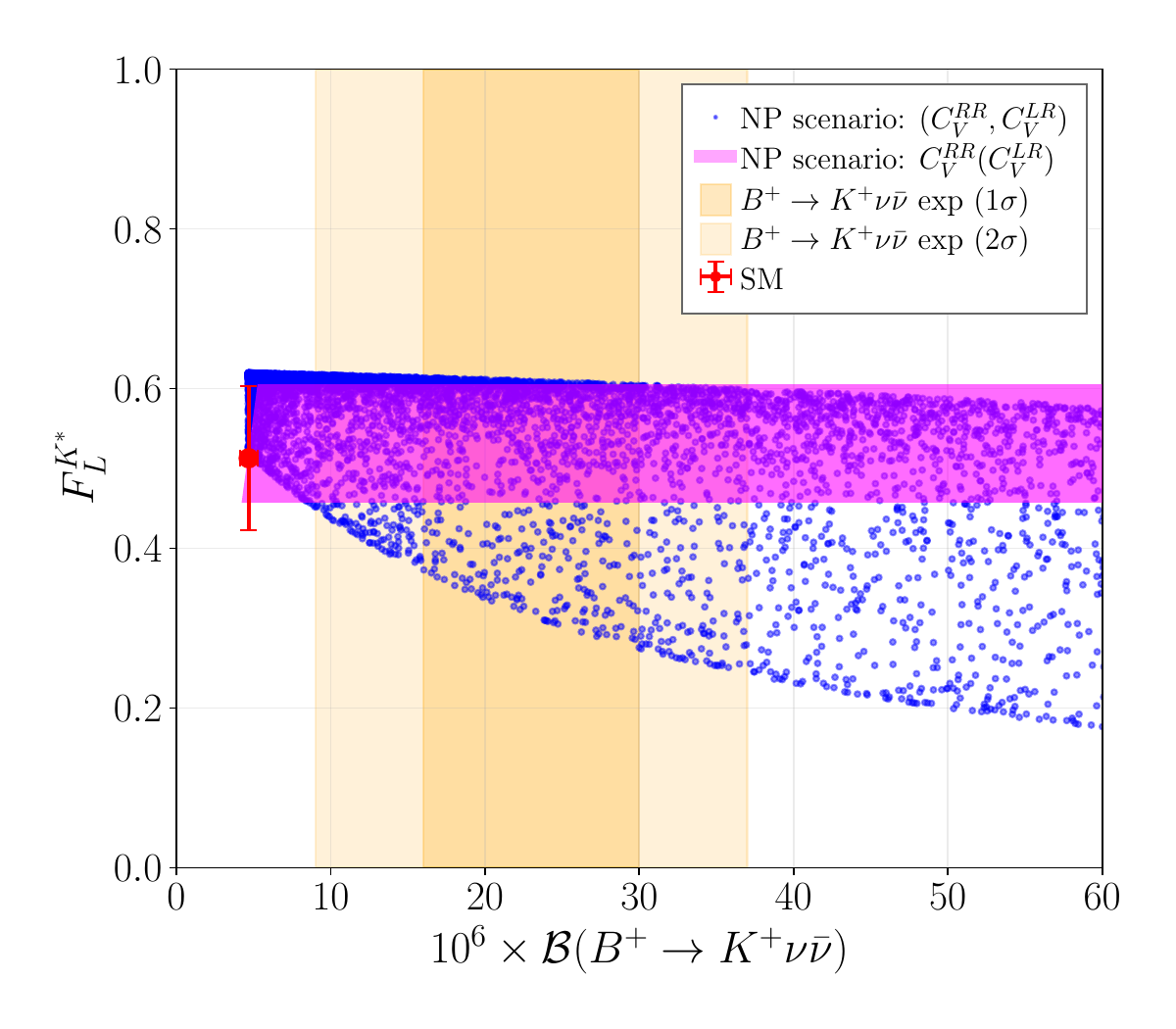}
\caption{Correlation of $\mathcal{B}(B \to K\nu\bar{\nu})$ with $\mathcal{B}(B \to K^*\nu\bar{\nu})$ (left) and $F_L^{K^\ast}$ (right) under RHN-scenario.}
\label{fig:BK_BKstr}
\end{figure}

The left panel of figure~\ref{fig:BK_BKstr} shows the correlation between $\mathcal{B} (B^+\to K^+\nu\bar{\nu})$ and $\mathcal{B}(B\to K^*\nu\bar{\nu})$ in the RHN scenario. The magenta band corresponds to varying a single Wilson coefficient ($\mathcal{C}_V^{RR}$ or $\mathcal{C}_V^{LR}$) at a time, while the blue points show the result of a simultaneous scan over both $(\mathcal{C}_V^{RR},\,\mathcal{C}_V^{LR})$. The SM prediction is shown as the red marker. The darker and lighter orange bands correspond to the $1\sigma$ and $2\sigma$ intervals of $\mathcal{B}(B^+\to K^+\nu\bar{\nu})$ from Belle~II, while the gray hatched region represents the current Belle upper limit on $\mathcal{B}(B\to K^*\nu\bar{\nu})$.

As seen from the figure, the RHN interactions shift both branching fractions away from their SM values, with the two observables exhibiting a strong positive correlation. In the single-coefficient scenario (magenta), the predicted values either fall below the Belle~II $1\sigma$ interval of $\mathcal{B}(B^+\to K^+\nu\bar{\nu})$ or exceed the Belle upper limit on $\mathcal{B}(B\to K^*\nu\bar{\nu})$, leaving no simultaneously consistent parameter space. The two-coefficient scenario (blue), on the other hand, yields a small consistent region where both constraints are satisfied. The upper limit on $\mathcal{B}(B\to K^*\nu\bar{\nu})$ proves to be particularly restrictive, eliminating the bulk of the two-coefficient parameter space and disfavoring scenarios that predict large $K^*$ rates even when consistent with the measured $\mathcal{B}(B^+\to K^+\nu\bar{\nu})$ at $1\sigma$ and $2\sigma$. This underscores the importance of future precise measurements of 
$\mathcal{B}(B\to K^*\nu\bar{\nu})$ at Belle~II and LHCb in further narrowing the allowed parameter space of right-handed neutrino interactions.

The right panel of figure~\ref{fig:BK_BKstr} shows the correlation between $\mathcal{B}(B^+\to K^+\nu\bar{\nu})$ and $F_L^{K^*}$ in the  RHN scenario. In the single-coefficient scenario (magenta), $F_L^{K^*}$ remains close to its SM value across the entire range of $\mathcal{B}(B^+\to K^+\nu\bar{\nu})$, which can be understood directly from equation~\ref{eq:dGamma_Kst}. With only one non-zero Wilson coefficient, the interference term vanishes identically, and the remaining NP longitudinal and transverse contributions scale almost similar with the single coefficient, leaving $F_L^{K^*}$ consistent with the SM. 

In the two-coefficient scenario (blue), $F_L^{K^*}$ develops a spread in both directions relative to the SM prediction. The downward deviations are driven by the transverse NP contribution, which grows with both WCs, while the upward deviations arise from the interference term, which is negative for same-sign coefficients, reducing the total decay width relative to the longitudinal width. The minimum of $F_L^{K^*}$ is reached when $\mathcal{C}_V^{LR} = \mathcal{C}_V^{RR}$, where the NP contribution to longitudinal decay rate vanishes. Within the experimentally allowed $1\sigma$ interval of 
$\mathcal{B}(B^+\to K^+\nu\bar{\nu})$, $F_L^{K^*}$ remains spread between approximately $0.3$ and $0.6$, demonstrating that a measurement of $F_L^{K^*}$ would provide additional constraints on the RHN operator structure, narrowing the allowed parameter space beyond what the branching fraction alone can achieve. Such a measurement is particularly promising at Belle~II and future FCC-ee.

\begin{figure}[!htb]
\centering
\includegraphics[width=0.49\textwidth]{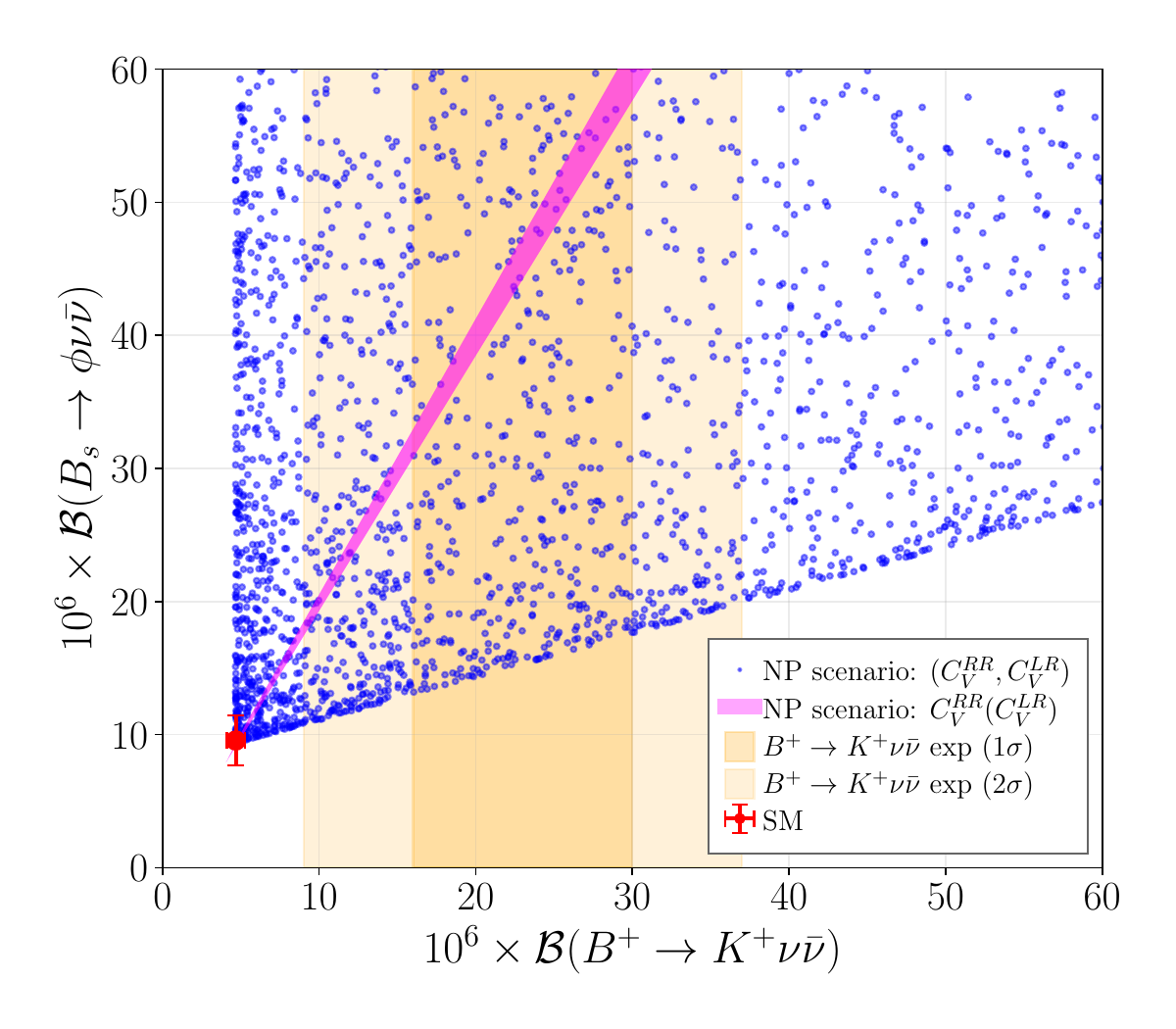}
\includegraphics[width=0.49\textwidth]{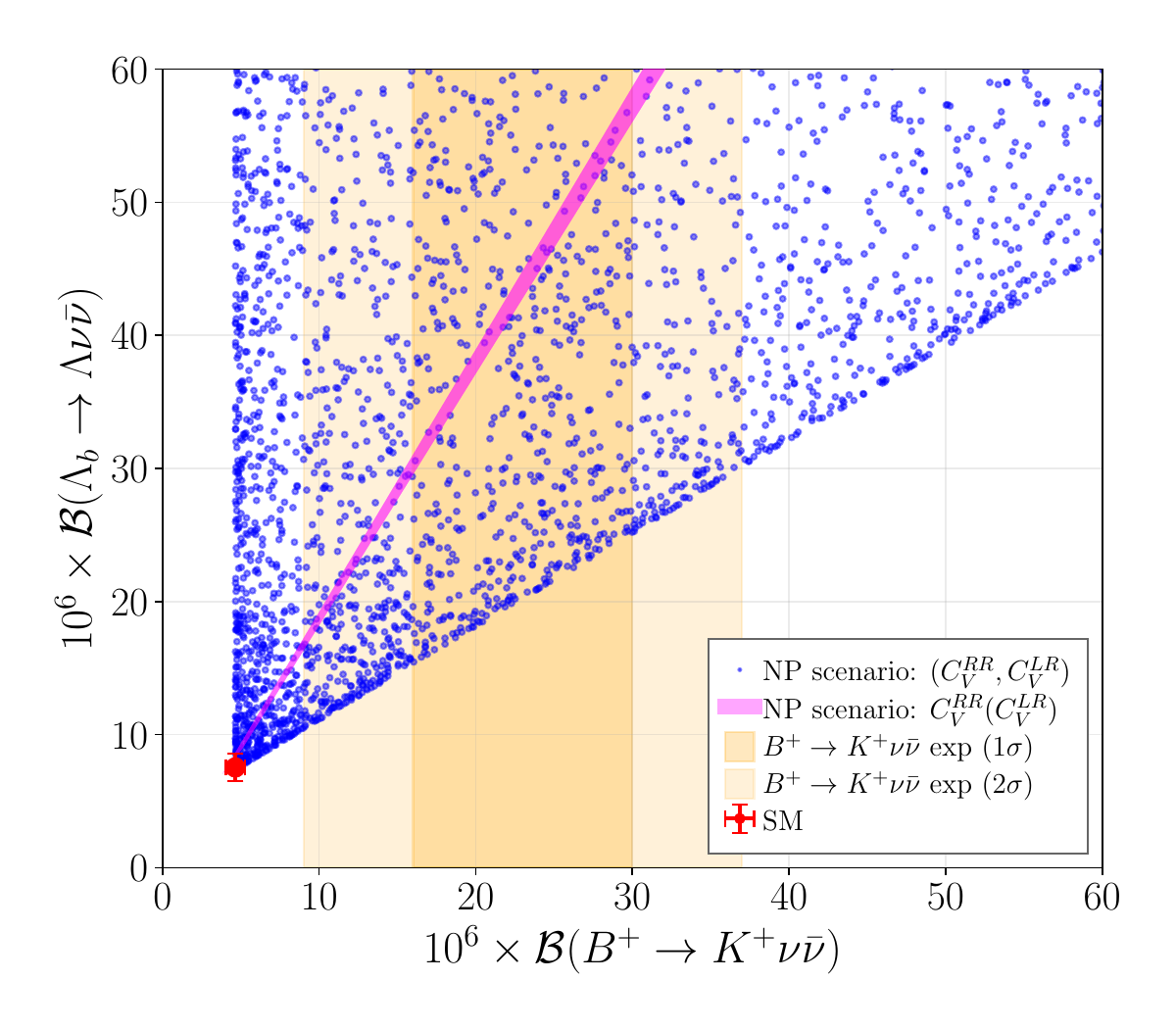}
\caption{Correlations of $\mathcal{B}(B \to K\nu\bar{\nu})$ with $\mathcal{B}(B_s \to \phi\nu\bar{\nu})$ (left) and $\mathcal{B}(\Lambda_b \to \Lambda\nu\bar{\nu})$ (right)  under RHN-scenario.}
\label{fig:BK_predictions}
\end{figure}
 
The left and right panels of figure~\ref{fig:BK_predictions} show the correlations of $\mathcal{B}(B^+\to K^+\nu\bar{\nu})$ with $\mathcal{B}(B_s\to\phi\nu\bar{\nu})$ and $\mathcal{B}(\Lambda_b\to\Lambda\nu\bar{\nu})$, respectively. In both cases, a strong positive correlation is observed, reflecting the fact that all three decays are governed by the same underlying $b\to s\nu\bar{\nu}$ transition and depend on the same WCs $\mathcal{C}_V^{LR}$ and $\mathcal{C}_V^{RR}$. In the single-coefficient scenario (magenta), the correlation is linear, since with only one non-zero Wilson coefficient all branching fractions scale as $|\mathcal{C}|^2$ and their ratios are fixed entirely by hadronic quantities. In the two-coefficient scenario (blue), the parameter space opens up considerably, generating a broad allowed region whose lower envelope reflects the minimum decay rate consistent with the underlying hadronic and kinematic structure. Any enhancement in $\mathcal{B}(B^+\to K^+\nu\bar{\nu})$ driven by the RHN interactions is necessarily accompanied by corresponding enhancements in $\mathcal{B}(B_s\to\phi\nu\bar{\nu})$ and $\mathcal{B}(\Lambda_b\to\Lambda\nu\bar{\nu})$. Since neither $\mathcal{B}(B_s\to\phi\nu\bar{\nu})$ nor 
$\mathcal{B}(\Lambda_b\to\Lambda\nu\bar{\nu})$ has been measured experimentally; these correlations constitute testable predictions of the RHN-scenario at LHCb and FCC-ee~\cite{FCC:2025lpp}.

\section{Conclusion} \label{sec:conclusion}

In this work, we have investigated the effects of right-handed neutrino interactions on rare $b\to s\nu\bar{\nu}$ transitions within a model-independent effective field theory framework. The two distinct 
allowed islands identified in the $(\mathcal{C}_V^{LR},\,\mathcal{C}_V^{RR})$ 
parameter space, consistent with both the Belle~II measurement of $\mathcal{B}(B^+\to K^+\nu\bar{\nu})$ and the Belle upper limit on $\mathcal{B}(B\to K^*\nu\bar{\nu})$, demonstrate that right-handed 
neutrino contributions can accommodate the observed excess while satisfying existing experimental constraints. The two allowed islands 
emerge from the simultaneous application of both constraints, neither of which alone is sufficient to isolate the viable parameter space, highlighting the importance of experimental studies of the $K^*$ mode in constraining the RHN operator structure.

The predicted correlated enhancements across $B\to K\nu\bar{\nu}$, $B\to K^*\nu\bar{\nu}$, $B_s\to\phi\nu\bar{\nu}$, and $\Lambda_b\to\Lambda\nu\bar{\nu}$ constitute a characteristic pattern of the RHN scenario, whose simultaneous observation across multiple decay modes would motivate further exploration of right-handed neutrino contributions to $b\to s\nu\bar{\nu}$ transitions. The longitudinal polarization 
fraction $F_L^{K^*}$ plays a distinctive and complementary role in this picture, unlike the branching fractions, it is sensitive to the 
relative sign and magnitude of $\mathcal{C}_V^{LR}$ and 
$\mathcal{C}_V^{RR}$, providing independent information on the operator structure beyond what the decay rate measurements alone can reveal. Since $\mathcal{B}(B_s\to\phi\nu\bar{\nu})$ and 
$\mathcal{B}(\Lambda_b\to\Lambda\nu\bar{\nu})$ remain experimentally unmeasured, the correlations predicted here represent testable forecasts of the RHN scenario. Precision measurements of $\mathcal{B}(B\to K^{(*)}\nu\bar{\nu})$ at Belle~II will narrow the 
allowed parameter space of $\mathcal{C}_V^{LR}$ and $\mathcal{C}_V^{RR}$, and together with future searches for $B_s\to\phi\nu\bar{\nu}$ and $\Lambda_b\to\Lambda\nu\bar{\nu}$ at LHCb and proposed FCC-ee, will constrain the right-handed neutrino interpretation of the observed anomaly in $b\to s\nu\bar{\nu}$ transitions.

\section*{Acknowledgements}
D.K. and J.K. acknowledge financial support from the Anusandhan National Research Foundation (ANRF) under grant no.~SERB/EEQ/2021/000965 and ~SERB/EEQ/2023/000959, respectively. Prisha and P.B. acknowledge the Ministry of Education (MoE), Government of India, for financial support.
\bibliographystyle{JHEPpri}
\bibliography{reference}

@article{LHCb:2022qnv,
    author = "Aaij, R. and others",
    collaboration = "LHCb",
    title = "{Test of lepton universality in $b \rightarrow s \ell^+ \ell^-$ decays}",
    eprint = "2212.09152",
    archivePrefix = "arXiv",
    primaryClass = "hep-ex",
    reportNumber = "LHCb-PAPER-2022-046, CERN-EP-2022-277",
    doi = "10.1103/PhysRevLett.131.051803",
    journal = "Phys. Rev. Lett.",
    volume = "131",
    number = "5",
    pages = "051803",
    year = "2023"
}

@article{LHCb:2014vgu,
    author = "Aaij, Roel and others",
    collaboration = "LHCb",
    title = "{Test of lepton universality using $B^{+}\rightarrow K^{+}\ell^{+}\ell^{-}$ decays}",
    eprint = "1406.6482",
    archivePrefix = "arXiv",
    primaryClass = "hep-ex",
    reportNumber = "CERN-PH-EP-2014-140, LHCB-PAPER-2014-024",
    doi = "10.1103/PhysRevLett.113.151601",
    journal = "Phys. Rev. Lett.",
    volume = "113",
    pages = "151601",
    year = "2014"
}

@article{LHCb:2017avl,
    author = "Aaij, R. and others",
    collaboration = "LHCb",
    title = "{Test of lepton universality with $B^{0} \rightarrow K^{*0}\ell^{+}\ell^{-}$ decays}",
    eprint = "1705.05802",
    archivePrefix = "arXiv",
    primaryClass = "hep-ex",
    reportNumber = "LHCB-PAPER-2017-013, CERN-EP-2017-100",
    doi = "10.1007/JHEP08(2017)055",
    journal = "JHEP",
    volume = "08",
    pages = "055",
    year = "2017"
}

@article{LHCb:2019hip,
    author = "Aaij, Roel and others",
    collaboration = "LHCb",
    title = "{Search for lepton-universality violation in $B^+\to K^+\ell^+\ell^-$ decays}",
    eprint = "1903.09252",
    archivePrefix = "arXiv",
    primaryClass = "hep-ex",
    reportNumber = "LHCb-PAPER-2019-009, CERN-EP-2019-043, LHCb-PAPER-2019-009 CERN-EP-2019-043",
    doi = "10.1103/PhysRevLett.122.191801",
    journal = "Phys. Rev. Lett.",
    volume = "122",
    number = "19",
    pages = "191801",
    year = "2019"
}

@article{LHCb:2019efc,
    author = "Aaij, Roel and others",
    collaboration = "LHCb",
    title = "{Test of lepton universality with $ {\Lambda}_b^0\to {pK}^{-}{\mathrm{\ell}}^{+}{\mathrm{\ell}}^{-} $ decays}",
    eprint = "1912.08139",
    archivePrefix = "arXiv",
    primaryClass = "hep-ex",
    reportNumber = "LHCb-PAPER-2019-040, CERN-EP-2019-272",
    doi = "10.1007/JHEP05(2020)040",
    journal = "JHEP",
    volume = "05",
    pages = "040",
    year = "2020"
}

@article{LHCb:2021trn,
    author = "Aaij, Roel and others",
    collaboration = "LHCb",
    title = "{Test of lepton universality in beauty-quark decays}",
    eprint = "2103.11769",
    archivePrefix = "arXiv",
    primaryClass = "hep-ex",
    reportNumber = "LHCb-PAPER-2021-004, CERN-EP-2021-042",
    doi = "10.1038/s41567-023-02095-3",
    journal = "Nature Phys.",
    volume = "18",
    number = "3",
    pages = "277--282",
    year = "2022",
    note = "[Addendum: Nature Phys. 19, (2023)]"
}

@article{LHCb:2021lvy,
    author = "Aaij, Roel and others",
    collaboration = "LHCb",
    title = "{Tests of lepton universality using $B^0\to K^0_S \ell^+ \ell^-$ and $B^+\to K^{*+} \ell^+ \ell^-$ decays}",
    eprint = "2110.09501",
    archivePrefix = "arXiv",
    primaryClass = "hep-ex",
    reportNumber = "LHCb-PAPER-2021-038, CERN-EP-2021-208",
    doi = "10.1103/PhysRevLett.128.191802",
    journal = "Phys. Rev. Lett.",
    volume = "128",
    number = "19",
    pages = "191802",
    year = "2022"
}

@article{Belle:2016fev,
    author = "Wehle, S. and others",
    collaboration = "Belle",
    title = "{Lepton-Flavor-Dependent Angular Analysis of $B\to K^\ast \ell^+\ell^-$}",
    eprint = "1612.05014",
    archivePrefix = "arXiv",
    primaryClass = "hep-ex",
    reportNumber = "BELLE-PREPRINT-2016-15, KEK-PREPRINT-2016-54",
    doi = "10.1103/PhysRevLett.118.111801",
    journal = "Phys. Rev. Lett.",
    volume = "118",
    number = "11",
    pages = "111801",
    year = "2017"
}

@article{BaBar:2012mrf,
    author = "Lees, J. P. and others",
    collaboration = "BaBar",
    title = "{Measurement of Branching Fractions and Rate Asymmetries in the Rare Decays $B \to K^{(*)} l^+ l^-$}",
    eprint = "1204.3933",
    archivePrefix = "arXiv",
    primaryClass = "hep-ex",
    reportNumber = "SLAC-PUB-14957, BABAR-PUB-12-002",
    doi = "10.1103/PhysRevD.86.032012",
    journal = "Phys. Rev. D",
    volume = "86",
    pages = "032012",
    year = "2012"
}

@article{BELLE:2019xld,
    author = "Choudhury, S. and others",
    collaboration = "BELLE",
    title = "{Test of lepton flavor universality and search for lepton flavor violation in $B \rightarrow K\ell \ell$ decays}",
    eprint = "1908.01848",
    archivePrefix = "arXiv",
    primaryClass = "hep-ex",
    reportNumber = "BELLE-CONF-1904, Belle Preprint 2020-11, KEK Preprint 2020-12",
    doi = "10.1007/JHEP03(2021)105",
    journal = "JHEP",
    volume = "03",
    pages = "105",
    year = "2021"
}

@article{Glashow:1970gm,
    author = "Glashow, S. L. and Iliopoulos, J. and Maiani, L.",
    title = "{Weak Interactions with Lepton-Hadron Symmetry}",
    doi = "10.1103/PhysRevD.2.1285",
    journal = "Phys. Rev. D",
    volume = "2",
    pages = "1285--1292",
    year = "1970"
}

@article{Crivellin:2025qsq,
    author = "Crivellin, Andreas and Iguro, Syuhei and Kitahara, Teppei",
    title = "{Discriminating tauphilic leptoquark explanations of the B anomalies via K{\textrightarrow}{\ensuremath{\pi}}{\ensuremath{\nu}}{\ensuremath{\nu}}{\textasciimacron} and B{\textrightarrow}K{\ensuremath{\nu}}{\ensuremath{\nu}}{\textasciimacron}}",
    eprint = "2505.05552",
    archivePrefix = "arXiv",
    primaryClass = "hep-ph",
    reportNumber = "ZU-TH 32/25, KEK-TH-2718, CHIBA-EP-271",
    doi = "10.1103/4dpx-h5vm",
    journal = "Phys. Rev. D",
    volume = "112",
    number = "9",
    pages = "095016",
    year = "2025"
}

@article{Chen:2025npb,
    author = "Chen, Chuan-Hung and Chiang, Cheng-Wei and de la Vega, Leon M. G.",
    title = "{Leptoquark-mediated Dirac neutrino mass and its impact on $ B\to K\nu \overline{\nu} $ and $ K\to \pi \nu \overline{\nu} $ decays}",
    eprint = "2503.22431",
    archivePrefix = "arXiv",
    primaryClass = "hep-ph",
    doi = "10.1007/JHEP09(2025)055",
    journal = "JHEP",
    volume = "09",
    pages = "055",
    year = "2025"
}

@article{Becirevic:2024iyi,
    author = "Be{\v{c}}irevi{\'c}, Damir and Fajfer, Svjetlana and Ko{\v{s}}nik, Nejc and Pavi{\v{c}}i{\'c}, Lovre",
    title = "{Right-handed interactions in puzzling B-decays}",
    eprint = "2410.23257",
    archivePrefix = "arXiv",
    primaryClass = "hep-ph",
    doi = "10.1016/j.physletb.2025.139285",
    journal = "Phys. Lett. B",
    volume = "861",
    pages = "139285",
    year = "2025"
}

@article{Browder:2021hbl,
    author = "Browder, Thomas E. and Deshpande, Nilendra G. and Mandal, Rusa and Sinha, Rahul",
    title = "{Impact of B{\textrightarrow}K{\ensuremath{\nu}}{\ensuremath{\nu}}{\textasciimacron} measurements on beyond the Standard Model theories}",
    eprint = "2107.01080",
    archivePrefix = "arXiv",
    primaryClass = "hep-ph",
    reportNumber = "SI-HEP-2021-19",
    doi = "10.1103/PhysRevD.104.053007",
    journal = "Phys. Rev. D",
    volume = "104",
    number = "5",
    pages = "053007",
    year = "2021"
}

@article{Altmannshofer:2023hkn,
    author = "Altmannshofer, Wolfgang and Crivellin, Andreas and Haigh, Huw and Inguglia, Gianluca and Martin Camalich, Jorge",
    title = "{Light new physics in B{\textrightarrow}K(*){\ensuremath{\nu}}{\ensuremath{\nu}}{\textasciimacron}?}",
    eprint = "2311.14629",
    archivePrefix = "arXiv",
    primaryClass = "hep-ph",
    reportNumber = "PSI-PR-23-46, ZU-TH 77/23",
    doi = "10.1103/PhysRevD.109.075008",
    journal = "Phys. Rev. D",
    volume = "109",
    number = "7",
    pages = "075008",
    year = "2024"
}

@article{Wang:2023trd,
    author = {Wang, Zeren Simon and Dreiner, Herbert K. and G{\"u}nther, Julian Y.},
    title = "{The decay $B\rightarrow K+\nu +\bar{\nu }$ at Belle II and a massless bino in R-parity-violating supersymmetry}",
    eprint = "2309.03727",
    archivePrefix = "arXiv",
    primaryClass = "hep-ph",
    doi = "10.1140/epjc/s10052-025-13745-6",
    journal = "Eur. Phys. J. C",
    volume = "85",
    number = "1",
    pages = "66",
    year = "2025"
}

@article{Aliev:2025hyp,
    author = "Aliev, T. M. and Elpe, A. and Selbuz, L. and Turan, I.",
    title = "{Explaining Belle data on B{\textrightarrow}K(*){\ensuremath{\nu}}{\ensuremath{\nu}}{\textasciimacron} decays via dark Z resonances}",
    eprint = "2503.22347",
    archivePrefix = "arXiv",
    primaryClass = "hep-ph",
    doi = "10.1103/6j6r-9vsl",
    journal = "Phys. Rev. D",
    volume = "112",
    number = "1",
    pages = "015025",
    year = "2025"
}

@article{Das:2025zrn,
    author = "Das, Diganta and Shameer, Dargi and Sain, Ria",
    title = "{$\Lambda_b \to \Lambda^{(\ast)}\nu\bar\nu$ decays and the recent Belle-II $B^+\to K^+\nu\bar\nu$ data}",
    eprint = "2507.01863",
    archivePrefix = "arXiv",
    primaryClass = "hep-ph",
    journal = "Phys. Rev. D",
    volume = "112",
    pages = "1",
    year = "2025"
}

@article{Escalona:2025jla,
    author = "Escalona, Patricio and Pinheiro, Jo{\~a}o Paulo and Oliveira, Vin{\'\i}cius and Doff, Adriano and De Sousa Pires, Carlos Antonio",
    title = "{Three Decades of FCNC Studies in 3-3-1 Model with Right-Handed Neutrinos: From Z'-Dominance to the Alignment Limit}",
    eprint = "2510.17979",
    archivePrefix = "arXiv",
    primaryClass = "hep-ph",
    doi = "10.3390/universe11120396",
    journal = "Universe",
    volume = "11",
    number = "12",
    pages = "396",
    year = "2025"
}

@article{Rosauro-Alcaraz:2024mvx,
    author = "Rosauro-Alcaraz, S. and Leal, L. P. S.",
    title = "{Disentangling left and right-handed neutrino effects in $B\rightarrow K^{(*)}\nu \nu $}",
    eprint = "2404.17440",
    archivePrefix = "arXiv",
    primaryClass = "hep-ph",
    doi = "10.1140/epjc/s10052-024-13104-x",
    journal = "Eur. Phys. J. C",
    volume = "84",
    number = "8",
    pages = "795",
    year = "2024"
}

@article{DelleRose:2019ukt,
    author = "Delle Rose, Luigi and Khalil, Shaaban and King, Simon J. D. and Moretti, Stefano",
    title = "{$R_K$ and $R_{K^*}$ in an Aligned 2HDM with Right-Handed Neutrinos}",
    eprint = "1903.11146",
    archivePrefix = "arXiv",
    primaryClass = "hep-ph",
    doi = "10.1103/PhysRevD.101.115009",
    journal = "Phys. Rev. D",
    volume = "101",
    number = "11",
    pages = "115009",
    year = "2020"
}

@article{Parrott:2022zte,
    author = "Parrott, W. G. and Bouchard, C. and Davies, C. T. H.",
    collaboration = "HPQCD",
    title = "{Standard Model predictions for B{\textrightarrow}K{\ensuremath{\ell}}+{\ensuremath{\ell}}-, B{\textrightarrow}K{\ensuremath{\ell}}1-{\ensuremath{\ell}}2+ and B{\textrightarrow}K{\ensuremath{\nu}}{\ensuremath{\nu}}{\textasciimacron} using form factors from Nf=2+1+1 lattice QCD}",
    eprint = "2207.13371",
    archivePrefix = "arXiv",
    primaryClass = "hep-ph",
    doi = "10.1103/PhysRevD.107.014511",
    journal = "Phys. Rev. D",
    volume = "107",
    number = "1",
    pages = "014511",
    year = "2023",
    note = "[Erratum: Phys.Rev.D 107, 119903 (2023)]"
}

@article{Belle-II:2023esi,
    author = "Adachi, I. and others",
    collaboration = "Belle-II",
    title = "{Evidence for B+{\textrightarrow}K+{\ensuremath{\nu}}{\ensuremath{\nu}}{\textasciimacron} decays}",
    eprint = "2311.14647",
    archivePrefix = "arXiv",
    primaryClass = "hep-ex",
    reportNumber = "Belle II Preprint 2023-017, KEK Preprint 2023-35",
    doi = "10.1103/PhysRevD.109.112006",
    journal = "Phys. Rev. D",
    volume = "109",
    number = "11",
    pages = "112006",
    year = "2024"
}

@article{Belle:2017oht,
    author = "Grygier, J. and others",
    collaboration = "Belle",
    title = "{Search for $\boldsymbol{B\to h\nu\bar{\nu}}$ decays with semileptonic tagging at Belle}",
    eprint = "1702.03224",
    archivePrefix = "arXiv",
    primaryClass = "hep-ex",
    doi = "10.1103/PhysRevD.96.091101",
    journal = "Phys. Rev. D",
    volume = "96",
    number = "9",
    pages = "091101",
    year = "2017",
    note = "[Addendum: Phys.Rev.D 97, 099902 (2018)]"
}

@article{Bailey:2015dka,
    author = "Bailey, Jon A. and others",
    title = "{$B\to Kl^+l^-$ Decay Form Factors from Three-Flavor Lattice QCD}",
    eprint = "1509.06235",
    archivePrefix = "arXiv",
    primaryClass = "hep-lat",
    reportNumber = "FERMILAB-PUB-15-403-T",
    doi = "10.1103/PhysRevD.93.025026",
    journal = "Phys. Rev. D",
    volume = "93",
    number = "2",
    pages = "025026",
    year = "2016"
}

@article{PhysRevD.110.030001,
  title = {Review of Particle Physics},
  author = {Navas, S. et. al,},
  collaboration = {Particle Data Group Collaboration},
  journal = {Phys. Rev. D},
  volume = {110},
  issue = {3},
  pages = {030001},
  numpages = {5},
  year = {2024},
  month = {Aug},
  publisher = {American Physical Society},
  doi = {10.1103/PhysRevD.110.030001},
  url = {https://link.aps.org/doi/10.1103/PhysRevD.110.030001}
}

@article{Bharucha:2015bzk,
    author = "Bharucha, Aoife and Straub, David M. and Zwicky, Roman",
    title = "{$B\to V\ell^+\ell^-$ in the Standard Model from light-cone sum rules}",
    eprint = "1503.05534",
    archivePrefix = "arXiv",
    primaryClass = "hep-ph",
    reportNumber = "TUM-HEP-957-14, CP3-Origins-2015-010, DIAS-2015-10",
    doi = "10.1007/JHEP08(2016)098",
    journal = "JHEP",
    volume = "08",
    pages = "098",
    year = "2016"
}

@article{Detmold:2016pkz,
    author = "Detmold, William and Meinel, Stefan",
    title = "{$\Lambda_b \to \Lambda \ell^+ \ell^-$ form factors, differential branching fraction, and angular observables from lattice QCD with relativistic $b$ quarks}",
    eprint = "1602.01399",
    archivePrefix = "arXiv",
    primaryClass = "hep-lat",
    reportNumber = "RBRC-1163, MIT-CTP-4767, RBRC-1163, MIT-CTP-4767",
    doi = "10.1103/PhysRevD.93.074501",
    journal = "Phys. Rev. D",
    volume = "93",
    number = "7",
    pages = "074501",
    year = "2016"
}

@article{Buras:2014fpa,
    author = "Buras, Andrzej J. and Girrbach-Noe, Jennifer and Niehoff, Christoph and Straub, David M.",
    title = "{$ B\to {K}^{\left(\ast \right)}\nu \overline{\nu} $ decays in the Standard Model and beyond}",
    eprint = "1409.4557",
    archivePrefix = "arXiv",
    primaryClass = "hep-ph",
    reportNumber = "FLAVOUR(267104)-ERC-80",
    doi = "10.1007/JHEP02(2015)184",
    journal = "JHEP",
    volume = "02",
    pages = "184",
    year = "2015"
}

@article{Belle-II:2018jsg,
    author = "Altmannshofer, W. and others",
    editor = "Kou, E. and Urquijo, P.",
    collaboration = "Belle-II",
    title = "{The Belle II Physics Book}",
    eprint = "1808.10567",
    archivePrefix = "arXiv",
    primaryClass = "hep-ex",
    reportNumber = "KEK Preprint 2018-27, BELLE2-PUB-PH-2018-001, FERMILAB-PUB-18-398-T, JLAB-THY-18-2780, INT-PUB-18-047, UWThPh 2018-26",
    doi = "10.1093/ptep/ptz106",
    journal = "PTEP",
    volume = "2019",
    number = "12",
    pages = "123C01",
    year = "2019",
    note = "[Erratum: PTEP 2020, 029201 (2020)]"
}

@article{Buchalla:1995vs,
    author = "Buchalla, Gerhard and Buras, Andrzej J. and Lautenbacher, Markus E.",
    title = "{Weak Decays beyond Leading Logarithms}",
    eprint = "hep-ph/9512380",
    archivePrefix = "arXiv",
    reportNumber = "SLAC-PUB-7009, SLAC-PUB-95-7009, MPI-PH-95-104, TUM-T31-100-95, FERMILAB-PUB-95-305-T",
    doi = "10.1103/RevModPhys.68.1125",
    journal = "Rev. Mod. Phys.",
    volume = "68",
    pages = "1125--1144",
    year = "1996"
}

@article{Altmannshofer:2021qrr,
    author = "Altmannshofer, Wolfgang and Stangl, Peter",
    title = "{New physics in rare B decays after Moriond 2021}",
    eprint = "2103.13370",
    archivePrefix = "arXiv",
    primaryClass = "hep-ph",
    doi = "10.1140/epjc/s10052-021-09725-1",
    journal = "Eur. Phys. J. C",
    volume = "81",
    number = "10",
    pages = "952",
    year = "2021"
}

@inbook{King:2025eqv,
    author = "King, Stephen F.",
    title = "{Right-handed neutrinos: seesaw models and signatures}",
    eprint = "2502.07877",
    archivePrefix = "arXiv",
    primaryClass = "hep-ph",
    month = "2",
    year = "2025"
}

@article{Albrecht:2021tul,
    author = "Albrecht, Johannes and van Dyk, Danny and Langenbruch, Christoph",
    title = "{Flavour anomalies in heavy quark decays}",
    eprint = "2107.04822",
    archivePrefix = "arXiv",
    primaryClass = "hep-ex",
    doi = "10.1016/j.ppnp.2021.103885",
    journal = "Prog. Part. Nucl. Phys.",
    volume = "120",
    pages = "103885",
    year = "2021"
}

@article{Buras:2009if,
    author = "Buras, Andrzej J.",
    title = "{Flavour Theory: 2009}",
    eprint = "0910.1032",
    archivePrefix = "arXiv",
    primaryClass = "hep-ph",
    doi = "10.22323/1.084.0024",
    journal = "PoS",
    volume = "EPS-HEP2009",
    pages = "024",
    year = "2009"
}

@article{Allwicher:2023xba,
    author = "Allwicher, Lukas and Becirevic, Damir and Piazza, Gioacchino and Rosauro-Alcaraz, Salvador and Sumensari, Olcyr",
    title = "{Understanding the first measurement of B(B{\textrightarrow}K{\ensuremath{\nu}}{\ensuremath{\nu}}{\textasciimacron})}",
    eprint = "2309.02246",
    archivePrefix = "arXiv",
    primaryClass = "hep-ph",
    doi = "10.1016/j.physletb.2023.138411",
    journal = "Phys. Lett. B",
    volume = "848",
    pages = "138411",
    year = "2024"
}

@article{Felkl:2021uxi,
    author = "Felkl, Tobias and Li, Sze Lok and Schmidt, Michael A.",
    title = "{A tale of invisibility: constraints on new physics in b {\textrightarrow} s{\ensuremath{\nu}}{\ensuremath{\nu}}}",
    eprint = "2111.04327",
    archivePrefix = "arXiv",
    primaryClass = "hep-ph",
    reportNumber = "CPPC-2021-13",
    doi = "10.1007/JHEP12(2021)118",
    journal = "JHEP",
    volume = "12",
    pages = "118",
    year = "2021"
}

@article{He:2021yoz,
    author = "He, Xiao Gang and Valencia, German",
    title = "{RK({\textasteriskcentered}){\ensuremath{\nu}} and non-standard neutrino interactions}",
    eprint = "2108.05033",
    archivePrefix = "arXiv",
    primaryClass = "hep-ph",
    doi = "10.1016/j.physletb.2021.136607",
    journal = "Phys. Lett. B",
    volume = "821",
    pages = "136607",
    year = "2021"
}

@article{Bause:2021cna,
    author = "Bause, Rigo and Gisbert, Hector and Golz, Marcel and Hiller, Gudrun",
    title = "{Interplay of dineutrino modes with semileptonic rare B-decays}",
    eprint = "2109.01675",
    archivePrefix = "arXiv",
    primaryClass = "hep-ph",
    reportNumber = "DO-TH 21/17",
    doi = "10.1007/JHEP12(2021)061",
    journal = "JHEP",
    volume = "12",
    pages = "061",
    year = "2021"
}

@article{Colangelo:1996ay,
    author = "Colangelo, P. and De Fazio, F. and Santorelli, Pietro and Scrimieri, E.",
    title = "{Rare $B \to K^{(*)}$ neutrino anti-neutrino decays at $B$ factories}",
    eprint = "hep-ph/9610297",
    archivePrefix = "arXiv",
    reportNumber = "BARI-TH-96-252, DSF-T-46-96, NAPOLI-PREPRINT-DSF-T-46-96",
    doi = "10.1016/S0370-2693(97)00130-5",
    journal = "Phys. Lett. B",
    volume = "395",
    pages = "339--344",
    year = "1997"
}

@article{Altmannshofer:2009ma,
    author = "Altmannshofer, Wolfgang and Buras, Andrzej J. and Straub, David M. and Wick, Michael",
    title = "{New strategies for New Physics search in $B \to K^{*} \nu \bar{\nu}$, $B \to K \nu \bar{\nu}$ and $B \to X_{s} \nu \bar{\nu}$ decays}",
    eprint = "0902.0160",
    archivePrefix = "arXiv",
    primaryClass = "hep-ph",
    reportNumber = "TUM-HEP-709-09",
    doi = "10.1088/1126-6708/2009/04/022",
    journal = "JHEP",
    volume = "04",
    pages = "022",
    year = "2009"
}

@article{Descotes-Genon:2020buf,
    author = "Descotes-Genon, S{\'e}bastien and Fajfer, Svjetlana and Kamenik, Jernej F. and Novoa-Brunet, Mart{\'\i}n",
    title = "{Implications of $b\to s\mu\mu$ anomalies for future measurements of $B \to K^{(*)} \nu \bar \nu$ and $K\to \pi \nu \bar \nu$}",
    eprint = "2005.03734",
    archivePrefix = "arXiv",
    primaryClass = "hep-ph",
    doi = "10.1016/j.physletb.2020.135769",
    journal = "Phys. Lett. B",
    volume = "809",
    pages = "135769",
    year = "2020",
    note = "[Addendum: Phys.Lett.B 840, 137830 (2023)]"
}

@article{Botella:2012ju,
    author = "Botella, F. J. and Branco, G. C. and Nebot, M.",
    title = "{The Hunt for New Physics in the Flavour Sector with up vector-like quarks}",
    eprint = "1207.4440",
    archivePrefix = "arXiv",
    primaryClass = "hep-ph",
    reportNumber = "IFIC-12-53",
    doi = "10.1007/JHEP12(2012)040",
    journal = "JHEP",
    volume = "12",
    pages = "040",
    year = "2012"
}

@article{Straub:2011gt,
    author = "Straub, David M.",
    editor = "Greco, Mario",
    title = "{New Physics Searches in Flavour Physics}",
    eprint = "1107.0266",
    archivePrefix = "arXiv",
    primaryClass = "hep-ph",
    doi = "10.1393/ncc/i2012-11132-x",
    journal = "Nuovo Cim. C",
    volume = "035N1",
    pages = "249--256",
    year = "2012"
}

@inproceedings{Crivellin:2016ivz,
    author = "Crivellin, Andreas",
    title = "{New Physics in the Flavour Sector}",
    booktitle = "{51st Rencontres de Moriond on QCD and High Energy Interactions}",
    eprint = "1605.02934",
    archivePrefix = "arXiv",
    primaryClass = "hep-ph",
    reportNumber = "PSI-PR-16-05",
    publisher = "ARISF",
    pages = "183--188",
    year = "2016"
}

@article{Fael:2025xmi,
    author = "Fael, Matteo and Jenkins, Jack and Lunghi, Enrico and Polonsky, Zachary",
    title = "{The Standard Model prediction for the rare decay $ B\to {X}_s\nu \overline{\nu} $}",
    eprint = "2512.19138",
    archivePrefix = "arXiv",
    primaryClass = "hep-ph",
    reportNumber = "P3H-25-113, SI-HEP-2025-32",
    doi = "10.1007/JHEP03(2026)217",
    journal = "JHEP",
    volume = "03",
    pages = "217",
    year = "2026"
}

@article{Hu:2025zua,
    author = "Hu, Quan-Yi and Duan, Zhi-Bin",
    title = "{Influence of invisible light particles on {\ensuremath{\Lambda}}b{\textrightarrow}{\ensuremath{\Lambda}}Emiss}",
    eprint = "2510.05272",
    archivePrefix = "arXiv",
    primaryClass = "hep-ph",
    doi = "10.1103/zp29-r1l6",
    journal = "Phys. Rev. D",
    volume = "112",
    number = "9",
    pages = "095007",
    year = "2025"
}

@article{Altmannshofer:2025eor,
    author = "Altmannshofer, Wolfgang and Gadam, Sri Aditya and Toner, Kevin",
    title = "{New strategies for new physics search with {\ensuremath{\Lambda}}b{\textrightarrow}{\ensuremath{\Lambda}}{\ensuremath{\nu}}{\ensuremath{\nu}}{\textasciimacron} decays}",
    eprint = "2501.10652",
    archivePrefix = "arXiv",
    primaryClass = "hep-ph",
    doi = "10.1103/PhysRevD.111.075005",
    journal = "Phys. Rev. D",
    volume = "111",
    number = "7",
    pages = "075005",
    year = "2025"
}

@article{Buras:2024mnq,
    author = "Buras, Andrzej J. and Stangl, Peter",
    title = "{On the interplay of constraints from $B_{s},$D,~ and K meson mixing in $Z^\prime $ models with implications for $b\!\rightarrow \! s \nu {\bar{\nu }}$ transitions}",
    eprint = "2412.14254",
    archivePrefix = "arXiv",
    primaryClass = "hep-ph",
    reportNumber = "AJB-24-3, CERN-TH-2024-218",
    doi = "10.1140/epjc/s10052-025-14168-z",
    journal = "Eur. Phys. J. C",
    volume = "85",
    number = "5",
    pages = "519",
    year = "2025"
}

@article{Felkl:2023ayn,
    author = "Felkl, Tobias and Giri, Anjan and Mohanta, Rukmani and Schmidt, Michael A.",
    title = "{When energy goes missing: new physics in $b\rightarrow s \nu \nu $ with sterile neutrinos}",
    eprint = "2309.02940",
    archivePrefix = "arXiv",
    primaryClass = "hep-ph",
    reportNumber = "CPPC-2023-06",
    doi = "10.1140/epjc/s10052-023-12326-9",
    journal = "Eur. Phys. J. C",
    volume = "83",
    number = "12",
    pages = "1135",
    year = "2023"
}

@article{FCC:2025lpp,
    author = "Benedikt, M. and others",
    collaboration = "FCC",
    title = "{Future Circular Collider Feasibility Study Report: Volume 1, Physics, Experiments, Detectors}",
    eprint = "2505.00272",
    archivePrefix = "arXiv",
    primaryClass = "hep-ex",
    reportNumber = "CERN-FCC-PHYS-2025-0002",
    doi = "10.1140/epjc/s10052-025-15077-x",
    journal = "Eur. Phys. J. C",
    volume = "85",
    number = "12",
    pages = "1468",
    year = "2025"
}
\end{document}